\documentclass[twocolumn,showpacs,amsmath,amssymb,aps]{revtex4}

\usepackage{graphicx}
\usepackage{dcolumn}
\usepackage{bm}
\usepackage{epsfig}

\begin{document}

\title{Coincidence detection of inharmonic pulses in a nonlinear crystal}

\author{Xavier Vidal$^{1,2}$}
\author{Pablo Balenzuela$^{3}$}
\email{balen@df.uba.ar}
\author{Javier M. Buld\'u$^{4,2}$}
\email{javier.martin-buldu@upc.edu}
\author{Jordi Martorell$^{1,2}$}
\author{Jordi Garc\'{\i}a-Ojalvo$^{2}$}
\affiliation{ $^{1}$ICFO - Institut de Ci\`encies Fot\`oniques,
Mediterranian Technology Park,
08860 Castelldefels, Spain \\
$^{2}$Departament de F\'{\i}sica i Enginyeria Nuclear, Universitat
Polit\`ecnica de Catalunya, Colom 11, E-08222 Terrassa, Spain\\
$^3$Departamento de F\'isica, Facultad de Ciencias Exactas y Naturales,
Universidad de Buenos Aires, Pabell\'on 1, Ciudad Universitaria (1428), Buenos Aires,
Argentina\\
$^4$Departamento de Ciencias de la Naturaleza y F\'{\i}sica Aplicada,
Universidad Rey Juan Carlos, Tulip\'an s/n,
28933 M\'ostoles, Spain.
}
\date{\today}

\begin{abstract}

Two trains of light pulses at periods that are equally shifted from the harmonics of a missing fundamental, are combined in a nonlinear crystal. As a result of a noncollinear phase matched second order nonlinear generation, a new train of pulses is obtained. When the temporal width of the input pulses is large, the frequency of the resulting pulse train follows the observations from classical experiments on the perception of virtual pitch by the brain. On the other hand, when the width of the input pulses is small, the generated pulse train exhibits much lower frequencies, analogous to those observed in the motor neural system. Our experimental set up allows us to explore, systematically and continuously, the transition between these two regimes, while at the same time demonstrate that coincidence detection in quadratic nonlinear systems has functionalities similar to those observed in the nervous system.

\end{abstract}

\maketitle



One of the most remarkable features of
nonlinear systems is their ability to process complex input signals.
A classical example is the perception of virtual pitch by the brain. In that
context, it is well known \cite{helmholtz85} that a missing fundamental tone
can be perceived upon exposure to only some of its harmonics. Psycophysical experiments by Schouten et al. \cite{schouten62} showed that
when a set of consecutive harmonics were equally shifted in frequency, it was not the
frequency difference (i.e. the original fundamental) that was perceived.
Instead, the perceived pitch varied
linearly with the frequency shift. Specifically, in the presence of input tones of
frequencies
\begin{eqnarray}
f_i=(k+i-1)f_0+\Delta f,\qquad i=1\ldots n
\label{eq:ghost_in}
\end{eqnarray}
where $f_0$ is the missing fundamental
frequency, $k>1$ is an integer
and $\Delta f$ is the frequency detuning which makes the input frequencies to be inharmonic.
Under these conditions, the perceived pitch was seen to
be given by a frequency that matches the following expression:
\begin{equation}
f_r=f_0+\frac{\Delta f}{k+\frac{(n-1)}{2}}
\label{eq:ghost_freq}
\end{equation}
Recently, Chialvo et al. \cite{chialvo02} proposed a simple and elegant mechanism
that accounts for this response, involving a linear superposition of the input
harmonics and a nonlinear noisy detection of the frequency (\ref{eq:ghost_freq}) via
a threshold. The mechanism, subsequently named {\em ghost resonance},
has been experimentally verified in lasers
\cite{buldu03,buldu05,sande05} and in electronic circuits \cite{calvo05}.

Experimental results using magnetoencephalographic measurements \cite{pantev96}
showed that the missing fundamental illusion also arises when the harmonic inputs are presented binaurally, i.e. different harmonics are applied to the
each of the two ears. The mechanism of Chialvo et al. \cite{chialvo02}
was extended
to that situation by modeling separate neuronal pathways that detected two
different input harmonics \cite{balenzuela05}. That study showed that in
the context of distributed inputs, the mechanism of ghost resonance heavily
relies of the coincidence detection of synaptic pulse trains (transduced by
the input neurons that receive the input harmonic signals) by an integrating
neuron. An experimental realization
of this effect in a real neurophysiological setup has been recently performed
\cite{manjarrez06}. That experiment has shown that when the input trains are
inharmonic (i.e. frequency shifted with respect to the original harmonics), the
processing neuronal pool responds at frequencies
much lower that those expected from expression (\ref{eq:ghost_freq}).
Subsequent experiments with nonlinear electronic circuits \cite{lopera06}
indicate that the difference in the response is due to the small width of the
pulses acting upon the integrating neuron, in contrast with the larger
width presumably associated with acoustic neuronal pathways.

In this paper, we use an utterly different experimental setup to address the
question of the influence of the input pulse width on the ghost resonance
response. This allows us to: (i) perform systematic measurements for
continuously varying pulse widths, with high controllability and reproducibility,
(ii) ascertain the generality of the phenomenon reported, which is seen
to arise in any system that operates via coincidence detection upon
thresholding, and (iii) suggest a possible functional role of this phenomenon
in nonlinear photonic devices, with potential applications in all-optical
signal processing.



\begin{figure}[htb]
\centering
\includegraphics[width=80mm,clip]{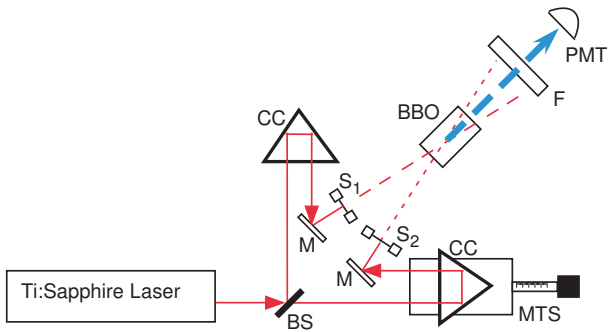}
\caption{Schematic representation of the experimental array. We
split the laser output, thin line, in two beams with a
beamsplitter, BS, and determine the frequencies of the pulses
using one shutter for each beam, S1 and S2, both of them
controlled by a code written in LabVIEW (National Instruments).
The thick dashed  line is for the second harmonic at the exit
of the nonlinear crystal, BBO; M are mirrors, F are filters to
stop the pump beam, CC are corner cubes and MTS indicates
micrometric translation stage. } \label{fig:fig01}
\end{figure}

The experimental setup is shown in Fig.~\ref{fig:fig01}. A 76~MHz
train of 130~fs pulses, produced by a Ti-sapphire laser at a
wavelength of 800~nm, was divided in two beams using a 50/50
beamsplitter. These two beams are recombined in a BBO crystal cut
for noncollinear phase matching, which generates second-harmonic
light at a wavelength of 400~nm when both beams are present
simultaneously in the crystal. The coincidence of the short laser
pulses is obtained using a movable translation stage, as shown in
Fig.~\ref{fig:fig01}. Both beams are chopped using
electro-mechanical shutters S$_1$ and S$_2$ at frequencies $f_1$
and $f_2$, respectively. The computerized chopping mechanism
allows to control the shutter frequencies and ensures a constant
(stable) phase relation between them.

\begin{figure}[htb]
\centering
\includegraphics[width=85mm,clip]{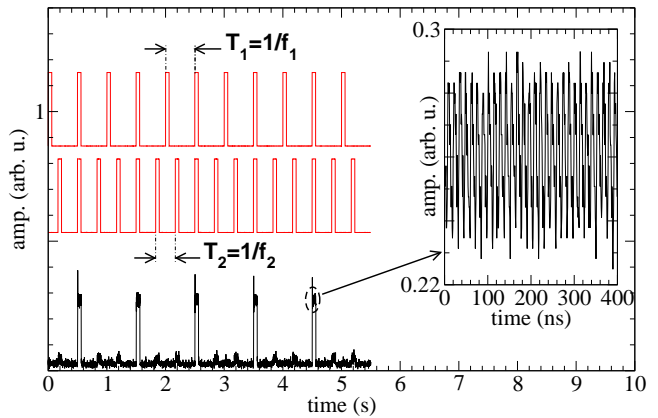}
\caption{ Time series of the input pulsed signals (two upper 
lines, shifted vertically for clarity) of frequencies $2$~Hz and
$3$~Hz, and of the response signal of the nonlinear crystal (lower
 line) at the ghost frequency $1$~Hz, as detected by the
photomultiplier. The inset shows the fully resolved time series of
a second-harmonic pulse, where the intensity does not go to zero
because of limited response time of the detection system.
Amplitude of the input signals has been vertically shifted in
order to ease comparison between time series. } \label{fig:fig02}
\end{figure}

The input signals consist on two pulse trains of dynamical
frequencies $f_1$ and $f_2$, represented by the two upper thin traces in
Fig.~\ref{fig:fig02}. In fact each pulse is, in turn, composed by
a package of ultrashort pulses with a repetition rate of $f_{\rm
rep}=76$~MHz and and optical frequency $f_{\rm opt}=3.75\times
10^{14}$~Hz, as explained above (see inset of Fig.
\ref{fig:fig02}). The width and dynamical frequency of each
package are controlled by the optical shutters. In what follows we
will refer to these packages as pulses, since the fast dynamics
within the package is not relevant for the purpose of our
experiment.

When two input pulses coincide inside the nonlinear crystal (see
Fig.~\ref{fig:fig02}), a train of second-harmonic pulses at an
optical frequency $2f_{\rm opt}$ is generated by the crystal, and
detected by the photomultiplier (lower  pulse train in
Fig.~\ref{fig:fig02}). Therefore, the nonlinear crystal acts as a
coincidence detector, and replaces the threshold of detection of
previous systems where ghost resonance has been studied
\cite{chialvo02,calvo05,balenzuela05,lopera06}. In the particular
case of Fig.~\ref{fig:fig02}, we have set $f_1=2$~Hz and
$f_2=3$~Hz, which corresponds to $n=2$, $k=2$, $f_0=1$~Hz and
$\Delta f=0$ in Eqs.~(\ref{eq:ghost_in})-(\ref{eq:ghost_freq}).
Given the thresholdless nature of the quadratic nonlinear
interaction, in this particular case a pulse train at the ghost
frequency ($f_r=1$~Hz [see Eq.~(\ref{eq:ghost_freq})]) is always
generated and its detection is only limited by the sensibility of
the overall detection system.

\begin{figure}[htb]
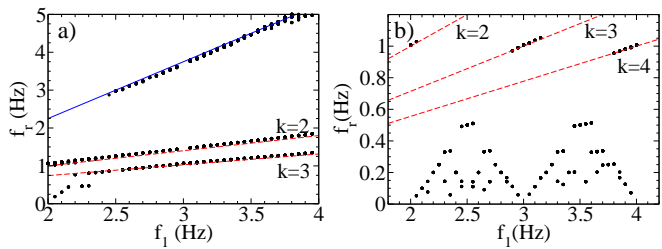

\centerline{
\includegraphics[width=42mm,clip]{fig03a.eps}
\includegraphics[width=43.5mm,clip]{fig03b.eps}}
\caption{ Inter-pulse instantaneous frequency for increasing
values of $\Delta f$. The horizontal axis corresponds to the slow
input frequency $f_1$ given by $f_1=kf_0+\Delta f$. The shutters
were here driven for trains of rectangular pulses of width $60$~ms
(a) and 5~ms (b). The dashed lines in panels (a) and  (b) indicate the 
expected ghost frequencies given by
Eq.~(\ref{eq:ghost_freq}), and the upper thin full line in panel
(a) represents the relation given in Eq.~(\ref{eq:blue}). }
\label{fig:fr}
\end{figure}

In order to investigate the response of the crystal to inharmonic
inputs and check the validity of Eq.~(\ref{eq:ghost_freq}), we set
the input frequencies to $f_1=kf_0+\Delta f$ and
$f_2=(k+1)f_0+\Delta f$, varying $\Delta f$ between 0 and 1~Hz.
Since we are concerned about the influence of the pulse width on
the pulse coincidence, we start by setting the pulse width $\Delta
t_p$ to a relatively large value, namely $\Delta t_p=60$~ms.
Figure~\ref{fig:fr}(a) shows the instantaneous response frequency
$f_r$ (defined as the inverse of the time interval between output
pulses) as a function of slowest input frequency $f_1$. The
response frequency is seen to follow well the relation predicted
by Eq.~(\ref{eq:ghost_freq}) for $k=2$ and $k=3$ (dashed
lines in the figure) for almost the whole range of $f_1$.
Lines of different $k$ are observed because
the input frequency $f_1$ can represent different harmonics of the fundamental 
frequency $f_0$, by changing the value of $k$. 

\begin{figure}[htb]
\centering
\includegraphics[width=70mm,clip]{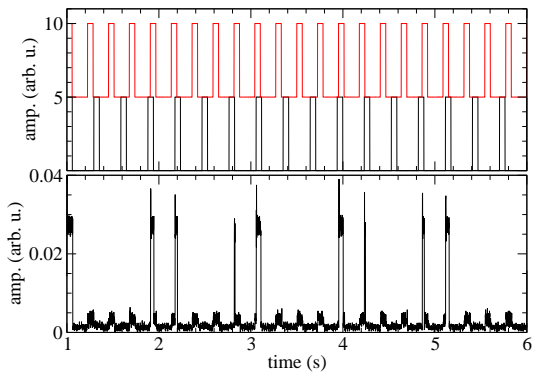}
\caption{
Time series of the input signals (upper panel, shifted vertically for clarity) and
response signal of the nonlinear crystal (lower panel), in the inharmonic case, for
frequencies $f_1=3.4$~Hz and $f_2=4.4$~Hz. We can see two dominant
frequencies in this response: the
lowest frequency corresponds to the values predicted by Eq.~(\ref{eq:ghost_freq}),
and the higher one arises from consecutive coincidences of the broad pulses
used in these sequence of experiments. Here the pulse width was set to $\Delta t_p=
60$~ms.
}
\label{fig:coinc}
\end{figure}

A linear response at large frequencies is also observed, however, in
plot~\ref{fig:fr}(a).
This high frequency response is a direct consequence of consecutive coincidences
of these broad pulses, as can be seen from Fig.~\ref{fig:coinc}. The resulting
frequency is the inverse of the silent period of the slower input signal
($T_1=\frac{1}{f_1}-\Delta t_p$), as can be deduced from this figure, and therefore,
for this case, the response frequency $f_r$ should follow the curve
\begin{equation}
f_r=\frac{1}{\frac{1}{f_1}-\Delta t_p}\,.
\label{eq:blue}
\end{equation}
This expression is represented by the upper thin solid line in
Fig.~\ref{fig:fr}(a), and exhibits good agreement with the
experimental results.

We now turn our attention to the opposite limit of very narrow pulses. To that
end we apply pulses of width $\Delta t_p=5$~ms (the narrowest we can get given
the speed limitations of the shutters). Figure~\ref{fig:fr}(b) shows the instantaneous response frequency $f_r$, again as a function of slowest input frequency, $f_1$.
We can see that the relation given by Eq.~(\ref{eq:ghost_freq}) holds only in the
neighborhood of the harmonic case ($f_1=2$~Hz and $3$~Hz, $f_r=1$~Hz).
On the other hand, in most of the inharmonic region, the system responds with
pulse trains of very low frequencies, in agreement with previous experimental
results in electronic circuits \cite{lopera06} and the motor neural system
\cite{manjarrez06}.
 These responses are grouped in families of lines following $f_r=\frac{\Delta f}{b}$, with $b$ being
an integer. The one with slope $1$ was observed experimentally in Ref. \cite{lopera06}
and its origin was determined analytically.
The closer we are to the limit of zero-width pulses, the more lines appear.
These lines arise from the condition of coincidence,
\begin{equation}
l~T_1=m~T_2,
\label{eq:t_coinc}
\end{equation}
where $l$, $m$ are integers and the $T_1$ and $T_2$ are the input periods,
respectively, $T_1=\frac{1}{f_1}=\frac{1}{kf_0+\Delta f}$
and $T_2=\frac{1}{f_2}=\frac{1}{(k+1)f_0+\Delta f}$. When the condition given by 
Eq. (\ref{eq:t_coinc})
is fulfilled, both input trains coincide and a pulse is detected.
When potential coincidences at frequency $\Delta f \neq 0$ are themselves missed, lower
frequencies $\frac{\Delta f}{b}$ are generated. This gives rise to different families
of lines [quasi-pyramids in Fig. \ref{fig:fr}-(b)], depending on the value of $k$
that relates $f_1$ and $f_0$.  

\begin{figure}[htb]
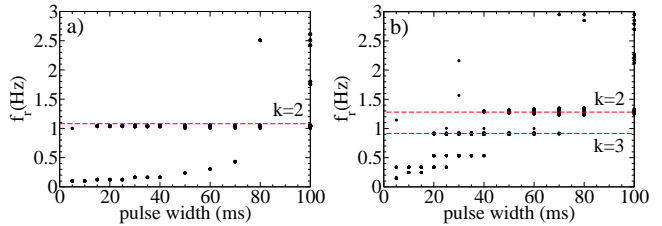

\centerline{
\includegraphics[width=42mm,clip]{fig05a.eps}
\includegraphics[width=42mm,clip]{fig05b.eps}}
\caption{
Inter-pulse instantaneous frequency for increasing values of $\Delta_f$  as a function of the pulse width. The input frequencies are (a) $f_1=2.1$~Hz and $f_2=3.1$~Hz,
and (b) $f_1=2.7$~Hz and $f_2=3.7$~Hz, which correspond respectively to a
detuning $\Delta f=0.1$~Hz (a) and $\Delta f=0.7$~Hz (b) with respect to the harmonic
$k=2$ frequencies $2.0$~Hz and $3.0$~Hz [see Eq.~(\ref{eq:ghost_freq})].
}
\label{fig:width}
\end{figure}

The two situations depicted in plots (a) and (b) of Fig.~\ref{fig:fr} represent
two opposite limits of the coincidence detection mechanism and show that,
in absence of a noisy threshold detection, 
the region of validity of Eq.~(\ref{eq:ghost_freq}) increases
with the pulse width.
In order to systematically investigate the transition between them, we now
fix the frequency shift $\Delta f$ and vary continuously the pulse width.
The corresponding result, for two different values of $\Delta f$, is shown in
Fig.~\ref{fig:width}.
For $\Delta f=0.1$~Hz, i.e. $f_1=2.1$~Hz and $f_2=3.1$~Hz (plot a), the $k=2$
line appears only for pulse widths larger than $15$~ms, and coexist with slow
frequency responses, which increase when the pulse width increases
\cite{lopera06}. The line corresponding to the case $k=3$ does not appear for this detuning.
For $\Delta f=0.7$~Hz, corresponding to
$f_1=2.7$~Hz and $f_2=3.7$~Hz (plot b), the response at $k=3$ appears
at pulse widths larger than $20$~ms, and the one at $k=2$ for widths larger
than $40$~ms. For broad enough pulses, the system responds at even
higher frequencies.

If we focus in the inharmonic region (when $\Delta f \ne 0$), the previous results clearly show the transition from a low-frequency response regime to another
regime completely dominated by relation 
(\ref{eq:ghost_freq}), as the width of the pulses increases. Such a transition between regimes
could underlie the differences observed  between previous psychophysical
experiments on auditory response \cite{schouten62} and recent experiments
on the motor neural system \cite{manjarrez06}. In the former, expression
(\ref{eq:ghost_freq}) held unambiguously; in the latter, on the other hand,
that behavior gave way very frequently to low frequency responses.
Correspondingly, synaptic pulses are known to be
wider in the auditory system than in the motor reflex system. From a
technological viewpoint, the present results also show that nonlinear
optical crystals carry out nontrivial signal processing tasks that mimic
those of more complicated systems such as the brain.

\begin{acknowledgments}
We thank Claudio Mirasso and El\'ias Manj\'arrez for fruitful
discussions. Financial support was provided by MCyT-FEDER (Spain,
projects BFM2003-07850, TEC2005-07799 and MAT2005-06354), by the
EC-funded project PHOREMOST (FP6/2003/IST/2511616), and by the
Generalitat de Catalunya.
\end{acknowledgments}

\end{document}